\documentclass[12pt, colorlinks=true, linkcolor=blue, urlcolor=blue, citecolor=blue]{article}

\usepackage[top=1in, bottom=1in, left=1in, right=1in]{geometry}

\usepackage[T1]{fontenc}
\usepackage{lmodern}

\usepackage{xcolor} 
\usepackage[colorlinks=true, linkcolor=blue, urlcolor=blue, citecolor=blue]{hyperref}
\usepackage{bookmark}

\usepackage{amsmath, amssymb, amsthm, mathtools, physics}

\usepackage{graphicx, subcaption, float, booktabs, multirow}
\usepackage{enumitem}
\setlist[enumerate]{nosep, topsep=0pt, leftmargin=*}

\usepackage{tikz}
\usetikzlibrary{calc, positioning, arrows.meta, shadows}
\tikzset{every picture/.append style={scale=1, transform shape}}

\usepackage{algorithm, algpseudocode}   
\makeatletter
\newcounter{breakablealgorithm}

\makeatother

\usepackage{natbib}
\usepackage{bibunits}
\defaultbibliographystyle{plainnat}  
\setcitestyle{round} 
\defaultbibliography{bibfile_main}

\newtheorem{theorem}{Theorem}[section]
\newtheorem{remark}{Remark}[section]

\newtheorem{proposition}{Proposition}[section]
\newtheorem{assumption}{Assumption}[section]

\usepackage{chngcntr}
\counterwithin{table}{section}
\counterwithin{algorithm}{section}
\counterwithin{figure}{section}
\numberwithin{equation}{section}

\allowdisplaybreaks[4]

\def\spacingset#1{\renewcommand{\baselinestretch}{#1}\small\normalsize}

\begin{document}

\spacingset{1.2}

\begin{bibunit}

\title{\Large Proximal Mediation Analysis with Hidden Recanting Witnesses}

\author{
    Sihan Wu \\
    Center for Data Science, Zhejiang University \\
    Yang Bai \\
    Department of Statistics and Data Science, National University of Singapore \\
    Yifan Cui\thanks{The authors gratefully acknowledge the financial support from the National Key R\&D Program of China (2024YFA1015600) and the National Natural Science Foundation of China (12471266 and U23A2064). Correspondence to \href{mailto:cuiyf@zju.edu.cn}{cuiyf@zju.edu.cn}.} \\
    Center for Data Science, Zhejiang University
}

\date{}
\maketitle

\begin{abstract}
Mediation analysis is essential for decomposing the causal effect of a treatment into direct and indirect pathways. However, many practical settings rely on the stringent assumption that recanting witnesses, defined as treatment-induced mediator-outcome confounders, are either absent or fully known a priori. Such a requirement is often untenable, especially when these variables remain unobservable due to measurement difficulties or privacy constraints. In this paper, we leverage proximal causal inference to develop three novel identification strategies to address the challenge of identifying path-specific effects in the presence of unknown recanting witnesses. Building on this, we develop a semiparametric inference framework that derives the efficient influence function and proposes a proximal multiply robust estimator, which remains consistent if at least one set of nuisance models is correctly specified. When all nuisance models are correctly specified and converge at appropriate rates, the estimator is asymptotically normal and achieves the semiparametric efficiency bound. We provide a minimax optimization-based debiased machine learning procedure for point estimation and constructing valid confidence intervals. The performance of the proposed methods is demonstrated by simulation studies and a real data application. 
\end{abstract}
	

\vfill
\newpage

\spacingset{2}

\newpage
\section{Introduction}

In recent years, causal mediation analysis \citep{Robins1992, Pearl2022, Vanderweele2009, Imai2010, TchetgenTchetgen2012} has gained substantial traction as a cornerstone of causal inference. Beyond evaluating the total treatment effect, the mediation framework allows researchers to disentangle the direct mechanism from the indirect pathway operating through an intermediate variable. Dissecting these distinct pathways with statistical rigor is essential for unraveling the underlying mechanisms that drive causal phenomena.

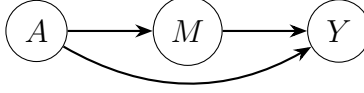
\begin{figure}[htbp]
\centering
\begin{tikzpicture}[
  >=Stealth,
  every node/.style={draw, circle, minimum size=8mm}
]

  \node (A) at (-2,0) {$A$}; 
  \node (M) at (0,0) {$M$};  
  \node (Y) at (2,0) {$Y$};

  \draw[->, thick] (A) -- (M);  
  \draw[->, thick, bend right=30] (A) to (Y);  
  \draw[->, thick] (M) -- (Y);  

\end{tikzpicture}
\caption{Causal diagram of a mediation model.}
\label{fig:Causal diagram of the mediation model}
\end{figure}

The simplest and most traditional mediation setting \citep{Robins1992, Pearl2022} involves a single intermediate variable $M$, as depicted in Figure~\ref{fig:Causal diagram of the mediation model}. 
Let $A$ denote the treatment variable with levels $A=0$ and $A=1$ (e.g., control and treated, respectively), $M$ denote the intermediate variable, and $Y$ denote the outcome variable. Following the potential outcomes model \citep{neyman1923applications,rubin1974estimating}, let $M(a)$ represent the potential value of the mediator under treatment level $a$, and $Y(m, a)$ represent the potential outcome if $M$ were set to $m$ and $A$ were set to $a$. Consequently, the average treatment effect can be decomposed as follows:
\begin{equation}
\label{equ: ATE}E[Y(1)-Y(0)] = \underbrace{E[Y(M(1), 1)-Y(M(0), 1)]}_{\text{Natural Indirect Effect (NIE)}} + \underbrace{E[Y(M(0), 1)-Y(M(0), 0)].}_{\text{Natural Direct Effect (NDE)}}
\end{equation}
Specifically, the Natural Indirect Effect (NIE) captures the expected outcome variation induced solely by changing the mediator from its potential value under $A=0$ to that under $A=1$, while freezing the treatment at $A=1$. Meanwhile, the Natural Direct Effect (NDE) isolates the expected change in the outcome driven by shifting the treatment from $0$ to $1$, while anchoring the mediator at its baseline counterfactual level $M(0)$.

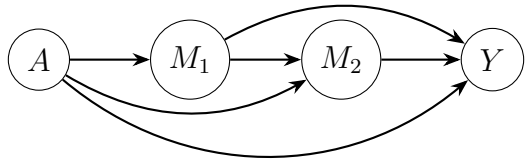
\begin{figure}[htbp]
\centering
\begin{tikzpicture}[
  >=Stealth,
  every node/.style={draw, circle, minimum size=8mm}
]

  \node (A)  at (-3,0) {$A$};
  \node (M1) at (-1,0) {$M_1$};
  \node (M2) at (1,0) {$M_2$};
  \node (Y)  at (3,0) {$Y$};

  \draw[->, thick] (A) -- (M1);
  \draw[->, thick] (M1) -- (M2);
  \draw[->, thick] (M2) -- (Y);

  \draw[->, thick, bend right=30] (A) to (M2);
  \draw[->, thick, bend right=40] (A) to (Y);

  \draw[->, thick, bend left=30] (M1) to (Y);

\end{tikzpicture}
\caption{Causal diagram of a mediation model with recanting witnesses.}
\label{fig:Causal diagram of recanting witness}
\end{figure}

In complex real-world systems, standard mediation analyses targeting NIE are often invalidated by the presence of so-called recanting witnesses \citep{Avin2005, Petersen2006}. To formally establish this structure, as illustrated in Figure~\ref{fig:Causal diagram of recanting witness}, we denote the primary mediator of interest as $M_2$ and the recanting witness as $M_1$. Structurally, the overall NIE represents the summation of the pure targeted effect ($A \to M_2 \to Y$) and the entangled sequential cascade ($A \to M_1 \to M_2 \to Y$). A fundamental identification deadlock arises here: adjusting for $M_1$ is necessary because it is a confounder between $M_2$ and $Y$, but conditioning on $M_1$ simultaneously blocks the sequential pathway $A \to M_1 \to M_2 \to Y$. Consequently, the NIE becomes unidentifiable. To bypass this structural bottleneck, the analytical focus must shift to the specific pathway where the treatment $A$ acts directly on $M_2$ to influence the outcome $Y$ ($A \to M_2 \to Y$). While this isolated path-specific effect remains identifiable, its estimation necessitates more advanced identification strategies.

Despite the conventional practice of directly assuming the complete absence of recanting witnesses, \cite{Miles2017, Miles2020} provide a rigorous framework for evaluating path-specific effects in the presence of recanting witnesses, and \cite{bai2026proximal} further extend this framework to accommodate pervasive unmeasured confounding. However, these methods require that researchers possess complete a priori knowledge of the true underlying structure of the recanting witnesses and that every constituent variable can be perfectly measured. In reality, researchers often lack definitive knowledge of the specific variables that actually comprise the recanting witnesses. Furthermore, even if the recanting witnesses could be fully enumerated, achieving complete measurement is often infeasible due to prohibitive data collection costs, ethical restrictions, or stringent privacy constraints. If any facet of the recanting witnesses remains obscure, these existing identification frameworks break down.

Recently, the proximal causal inference framework \citep{Miao2018, cui2024semiparametric}, as a novel tool to resolve unmeasured confounding in observational studies, has rapidly expanded to accommodate increasingly complex data structures. Specifically, utilizing this methodology, \cite{Tchetgen2020, Ying2023} address time-varying confounding in longitudinal studies; \cite{Ying2022, Ying2024} handle censored time-to-event outcomes in confounded survival analysis; \cite{shi2026theory} propose proximal synthetic control methods to settings with poor pre-treatment fit. Extending this logic to the mediation framework, \cite{Dukes2023} identify NIE and NDE in the presence of unmeasured confounding. Going a step further to bridge causal inference and reinforcement learning, researchers adapt the proximal framework to more general decision-making problems. \cite{qi2024proximal}, \cite{Shen2023} and \cite{sverdrup2023proximal} incorporate the proxy methods into the estimation of optimal individualized treatment regimes or heterogeneous treatment effects, while \cite{shi2022minimax, Zhang2026, Gao2025} apply it to dynamic treatment regimes under unmeasured confounding. Furthermore, \cite{wang2026blessing} introduce a super policy learning framework that leverages human-AI interaction to achieve a stronger decision oracle in confounded environments.

Beyond adjusting for latent confounders, the inherent flexibility of this proxy-based framework offers a pathway to tackle extreme structural missingness where key nodes in the causal graph are entirely hidden. For instance, recent studies apply the proximal inference framework to capture indirect pathways through hidden mediators \citep{Ghassami2025}, 
identify causal effects under hidden treatments \citep{zhou2024causal},  and construct robust estimators for settings where the outcome is hidden \citep{guo2026proximal}.

In this paper, we leverage the proximal causal inference framework to develop a new identification framework for path-specific effects when the recanting witnesses are hidden. Specifically, we establish novel nonparametric identification strategies and further advance the semiparametric inference theory by deriving the efficient influence function and constructing a multiply robust and locally efficient estimator. Moreover, we develop a practical estimation procedure based on a minimax optimization-based approach, formally proving its convergence rates and asymptotic efficiency. Critically, because our framework targets a hidden node rather than a confounder, the explicit formulations, especially the constructions of bridge functions and the subsequent identification arguments, differ in important ways from standard proximal causal inference paradigms. To the best of our knowledge, this is the first work to formalize and resolve the challenge of hidden recanting witnesses. By doing so, we not only provide a practical solution to a pervasive bottleneck in mediation analysis but also extend the applicability of proximal causal inference to complex graphical structures.

In Section~\ref{sec:Preliminaries}, we introduce necessary notation and assumptions. In Section~\ref{sec:identification}, we present three identification strategies via bridge functions. Section~\ref{sec:Semiparametric inference} advances the semiparametric inference framework by deriving the efficient influence function and proposing an efficient estimator. We then analyze the convergence rates and asymptotic normality of the proposed estimators. Furthermore, we detail a minimax optimization-based estimation procedure and propose corresponding statistical inference. Section~\ref{sec:simulation} reports simulation results that demonstrate the performance of our methods. Section~\ref{sec:real_data} applies the framework to a real-world study. The article concludes with a discussion of future work in Section~\ref{sec:disc}. Details of the proofs can be found in the Supplementary Material.

\section{Preliminaries}
\label{sec:Preliminaries}
\subsection{Notation}
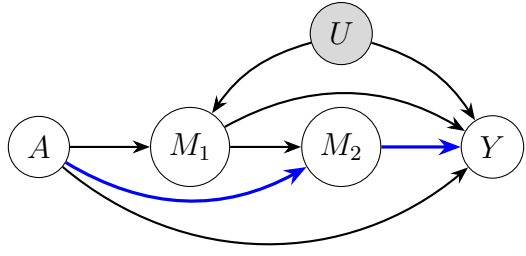
\begin{figure}[htbp]
\centering
\begin{tikzpicture}[
  >=Stealth,
  every node/.style={draw, circle, minimum size=6mm}
]

\node [fill=gray!30](U)  at (1,1.5) {$U$};
\node (A)  at (-3,0) {$A$};
\node (M1) at (-1,0) {$M_1$};
\node (M2) at (1,0) {$M_2$};
\node (Y)  at (3,0) {$Y$};

\draw[->, thick, bend right=20] (U) to (M1);
\draw[->, thick, bend left=20] (U) to (Y);

\draw[->, thick] (A) -- (M1);
\draw[->, very thick, bend right=30,blue] (A) to (M2);
\draw[->, thick, bend right=40] (A) to (Y);
%

\draw[->, thick] (M1) -- (M2);
\draw[->, thick, bend left=30] (M1) to (Y);
\draw[->, very thick, blue] (M2) -- (Y);

\end{tikzpicture}
\caption{Causal diagram of the sequential mediation model with unobserved confounding.}
\label{fig:Causal diagram of the sequential mediation model with U}
\end{figure}

In this section, we review the identification of the path-specific effect when recanting witnesses are fully observed \citep{Miles2017, Miles2020}. The corresponding causal graph is illustrated in Figure~\ref{fig:Causal diagram of the sequential mediation model with U}. The exposure $A$ directly affects the recanting witness $M_1$, the subsequent mediator $M_2$, and the outcome $Y$; simultaneously, $M_1$ directly affects both $M_2$ and $Y$, while $M_2$ also directly affects $Y$. Note that apart from the binary $A$, all other variables may be discrete, continuous, or multi-dimensional. 
Within this causal structure, an unmeasured confounder $U$ between $M_1$ and $Y$ is allowed to exist. 
The blue path is the causal pathway of interest that we will introduce in the next subsection.
To maintain visual clarity, we omit the baseline covariates $X$ that can directly influence all the aforementioned variables.

 Let $M_1(a)$ denote the potential value of the recanting witness if the treatment were set to $A=a$, and let $M_2(m_1, a)$ denote the potential value of the subsequent mediator if $M_1$ were set to $m_1$ and $A$ were set to $a$. This formulation allows us to construct nested potential outcomes. For example, $M_2\left(M_1(a_2), a_1\right)$ represents the potential value of $M_2$ if $A$ were set to $a_1$, with $M_1$ naturally taking its potential value under $A=a_2$. Similarly, let $Y(m_2, m_1, a)$ denote the potential outcome when $M_2=m_2$, $M_1=m_1$, and $A=a$. Building upon this definition, the nested counterfactual $Y(M_2(M_1(a_2), a_1), M_1(a_2), a_2)$ represents the potential outcome of $Y$ if the exposure $A$ were set to $a_2$, except for its direct effect on $M_2$, where it would instead be set to $a_1$. For simplicity, the natural potential outcome under a uniform treatment level $a \in \{0,1\}$ is defined as $Y(a) = Y(M_2(M_1(a), a), M_1(a), a)$.

\begin{remark}
It is crucial to emphasize that the two instances of $M_1(a_2)$ in the nested counterfactual $Y(M_2(M_1(a_2), a_1), M_1(a_2), a_2)$ represent the identical realized value for a given individual, rather than independent draws from its marginal distribution. Treating them as independent draws from the same distribution would sever the intra-individual correlation structure, failing to capture the shared individual-level characteristics.
\end{remark}

\subsection{Identification with an observed recanting witness}
We aim to identify the path-specific effect along the pathway $A \rightarrow M_2 \rightarrow Y$, which we denote as $P_{AM_2Y}$ \citep{Avin2005, Shpitser2013}. This represents the effect of $A$ on $Y$ that operates solely and directly through $M_2$. In Figure~\ref{fig:Causal diagram of the sequential mediation model with U}, this pathway is highlighted in blue. Using the nested counterfactual notation defined above, the path-specific effect of $A$ on $Y$ along the path  $A \rightarrow M_2 \rightarrow Y$ is formally written as
\begin{equation}
P_{AM_2Y}=E\left[ Y\big(M_2(M_1(0), 1),\, M_1(0),\, 0\big) - Y\big(M_2(M_1(0), 0),\, M_1(0),\, 0\big) \right].
\label{eq:formula1}
\end{equation}

Throughout, we implicitly assume standard consistency, meaning that the observed variables correspond to the potential outcomes under the observed treatment and mediator values: $M_1 = M_1(A)$, $M_2 = M_2(M_1, A)$, and $Y = Y(M_2, M_1, A)$. To identify the path-specific effect, \cite{Miles2020} posit the following three identifying assumptions.
\begin{assumption}[Positivity/Boundedness]
\label{asm:positivity}
For all values of $m_2, m_1$, $a$, and $x$ in their respective supports, we have
\begin{enumerate}
     \item[(i)] $0 < P(A = a \mid X = x) < 1. $
     \item[(ii)] $ 0< p(M_2 = m_2, M_1 = m_1 \mid A = a, X = x) < \infty. $
\end{enumerate}
\end{assumption}
\begin{assumption}[Sequential Ignorability]
\label{asm:seq_ign}
For all values $m_2$, $m_1$, and $a$ in their respective supports, the following conditional independence statements hold:
\begin{enumerate}
    \item[(i)] $\{Y(m_2, a), M_1(a)\} \perp\!\!\!\perp A \mid X$,
    \item[(ii)] $Y(m_2) \perp\!\!\!\perp M_2 \mid \{M_1, A, X\}$,
    \item[(iii)] $M_2(m_1, a) \perp\!\!\!\perp \{M_1, A\} \mid X$.
\end{enumerate}
\end{assumption}
\begin{assumption}[Cross-world Independence]
\label{asm:Cross-world Independence}
For all values $m_2$, $m_1$, $a$, and $a^{\prime}$, in their respective supports, the potential outcomes satisfy:
\[
\{Y(m_2, a), M_1(a)\} \perp\!\!\!\perp M_2(m_1, a^{\prime}) \mid X.
\]
\end{assumption}
Here, $Y(m_2, a)$ and $Y(m_2)$ serve as shorthand notations for the potential outcomes $Y(m_2, M_1(a), a)$ and $Y(m_2, M_1, A)$, respectively. 

Assumption~\ref{asm:positivity} is a standard positivity and boundedness requirement on the treatment assignment probability and density of mediators, respectively. Assumption~\ref{asm:seq_ign} enforces unconfoundedness among specific pairs of nodes. However, it is worth noting that the presence of unmeasured confounders between $M_1$ and $Y$ does not violate this assumption. Furthermore, by assuming the independence of specific variables across different counterfactual worlds, Assumption~\ref{asm:Cross-world Independence} makes the identification of nested counterfactuals possible.

Under these assumptions, the second term of Equation~\eqref{eq:formula1} simplifies to $E[Y(0)]$, representing the average outcome had all patients been assigned the baseline treatment level $A=0$. Under Assumption~\ref{asm:positivity}(i) and $Y(a) \perp\!\!\!\perp A \mid X$ implied by Assumption~\ref{asm:seq_ign}, its estimation has been extensively studied in the causal inference literature \citep{Robins1992}. Therefore, we focus on the first term of Equation~\eqref{eq:formula1} denoted by
\[
\psi = E\left[Y(M_2(M_1(0), 1), M_1(0), 0)\right].
\]
Building on the framework developed in \cite{Miles2017, Miles2020}, the expression for $\psi$ is identified under Assumptions~\ref{asm:positivity}-\ref{asm:Cross-world Independence},
\begin{equation}
\psi = \iiiint y  p(y \mid m_2, m_1, A=0, x)  p(m_2 \mid m_1, A=1, x)  p(m_1 \mid A=0, x)  p(x)  dy  dm_2 dm_1 dx.
\label{eq:formula2}
\end{equation}
Equation~\eqref{eq:formula2} provides a nonparametric identifying formula for the target parameter $\psi$ through a sequential integration over observable distributions.

\section{Identification with hidden recanting witnesses
}
\label{sec:identification}
\subsection{Outcome-model-based identification}

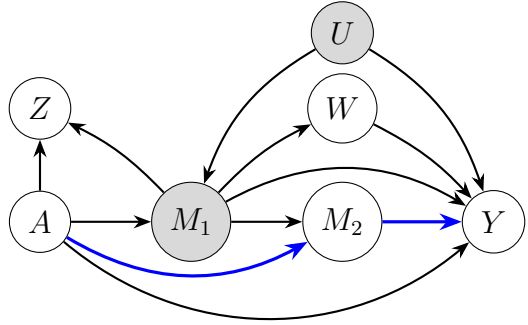
\begin{figure}[htbp]
\centering
\begin{tikzpicture}[
  >=Stealth,
  every node/.style={draw, circle, minimum size=8mm}
]

\node [fill=gray!30](U)  at (1,2.5) {$U$};
\node (A)  at (-3,0) {$A$};
\node [fill=gray!30](M1) at (-1,0) {$M_1$};
\node  (M2) at (1,0) {$M_2$};
\node (Y)  at (3,0) {$Y$};
\node (Z)  at (-3,1.5) {$Z$};
\node (W)  at (1,1.5) {$W$};

\draw[->, thick, bend right=10] (M1) to (Z);
\draw[->, thick, bend left=10] (M1) to (W);
\draw[->, thick] (A) -- (M1);
\draw[->, very thick, bend right=30,blue] (A) to (M2);
\draw[->, thick, bend right=40] (A) to (Y);
\draw[->, thick, bend left=20] (U) to (Y);
\draw[->, thick, bend right=20] (U) to (M1);
\draw[->, thick] (M1) -- (M2);
\draw[->, thick, bend left=30] (M1) to (Y);
\draw[->, very thick, blue] (M2) -- (Y);
\draw[->, thick, bend left=0] (A) to (Z);
\draw[->, thick, bend left=10] (W) to (Y);
\end{tikzpicture}
\caption{Causal diagram of the sequential mediation model with proxies.}
\label{fig:Causal diagram of the sequential mediation model with ZW}
\end{figure}

In our considered setting, the recanting witness $M_1$ is hidden, precluding direct access to its observed values. We introduce two proxy variables, $W$ and $Z$, which satisfy the assumptions below. The causal structure is depicted in Figure~\ref{fig:Causal diagram of the sequential mediation model with ZW}. Similar to the previous figure, for visual clarity, we have omitted the representation of the baseline covariates $X$. 

\begin{assumption}[Proxy Variables for $M_1$]
\label{asm:proxies}
There exist two observed proxy variables $Z$ and $W$ for the unobserved recanting witness $M_1$ such that:
\begin{enumerate}
    \item[(i)] $Z \perp\!\!\!\perp \{Y,M_2\} \mid \{M_1, A, X\}$;
    \item[(ii)] $W \perp\!\!\!\perp \{A, Z,M_2\} \mid \{M_1, X\}$.
\end{enumerate}
\end{assumption}
In addition, we assume that $0<p(W \mid A=a,X)<\infty$ for $a=0,1$ and $0<P(A=1 \mid W,M_2,X)<1$ almost surely hereinafter.

\begin{remark}
It is worth noting that Assumption~\ref{asm:proxies} accommodates both the presence and absence of the direct causal effects $A \rightarrow Z$ and $W \rightarrow Y$. Furthermore, this highlights a key structural distinction between our framework and conventional proximal causal inference \citep{cui2024semiparametric}. In standard proximal inference, the proxy $Z$ is often assumed to directly affect the treatment ($Z \rightarrow A$). However, in our sequential mediation setting, assuming $Z \rightarrow A$ would induce a causal cycle ($Z \rightarrow A \rightarrow M_1 \rightarrow Z$), which violates the fundamental property of a directed acyclic graph.
\end{remark}

\begin{assumption}[Completeness of $Z$ for $M_1$]
\label{asm:completeness}
\leavevmode 
\begin{enumerate}[label=(\roman*), itemsep=0.5ex, topsep=0.5ex]
    \item For any square-integrable function $g$ and any value $x$, if
    \[
        E[g(M_1) \mid Z, A = 1, X = x] = 0 \quad \text{almost surely},
    \]
    then $g(M_1) = 0$ almost surely.
    \item For any square-integrable function $g$ and any values $x, m_2$, if
    \[
        E[g(M_1) \mid Z, M_2 = m_2, A = 0, X = x] = 0 \quad \text{almost surely},
    \]
    then $g(M_1) = 0$ almost surely.
\end{enumerate}
\end{assumption}

\begin{assumption}[Bridge Function h]
\label{asm: bridge function 1}
Suppose that there exist confounding bridge functions $h_0(W,M_2,X)$ and
$h_1(W,X)$ that almost surely satisfy
\begin{align}
&E[Y \mid Z, M_2, A=0, X]=\int h_0(w,M_2,X)dF(w \mid Z, M_2, A=0, X),\label{eq:formulah1}\\
&E[h_0(W,M_2,X) \mid Z, A=1, X]=\int h_1(w,X)dF(w \mid Z, A=1, X).\label{eq:formulah2}
\end{align}
\end{assumption}

Assumption~\ref{asm:completeness} is commonly referred to as the completeness condition \citep{Newey2003}, which conceptually implies that $Z$ sufficiently captures the features of $M_1$. As noted by \cite{Tchetgen2020}, this requirement is naturally satisfied in discrete settings provided that $\min\{ \dim(Z), \dim(W) \} \geq \dim(M_1)$.
Mathematically, each of the integral equations in Assumption~\ref{asm: bridge function 1} defines an inverse problem known as a Fredholm integral equation of the first kind, and Assumption~\ref{asm: bridge function 1} ensures that this inverse problem admits a square-integrable solution, which is common practice in the proximal causal inference literature \citep{Miao2018, cui2024semiparametric}.

Under the assumptions above, we obtain the following proposition.

\begin{proposition}
\label{prop: bridge function 1}
By Assumptions~\ref{asm:positivity}-\ref{asm:Cross-world Independence}, and \ref{asm:proxies}-\ref{asm: bridge function 1}, we have the following two equations hold almost surely:
\begin{align}
&E[Y \mid M_2, M_1, A=0, X]=\int h_0(w,M_2,X)dF(w \mid M_2, M_1, A=0, X),
\label{eq:formula3}\\
&E[h_0(W,M_2,X) \mid M_1, A=1, X]=\int h_1(w,X)dF(w \mid M_1, A=1, X).
\label{eq:formula4}
\end{align}
\end{proposition}
Furthermore, building upon the above result, we propose our first identification strategy, named the Proximal Outcome Regression (POR) method.

\begin{theorem}[POR of $\psi$]
\label{thm: Identification 1}
Under Assumptions~\ref{asm:positivity}-\ref{asm:Cross-world Independence} and \ref{asm:proxies}-\ref{asm: bridge function 1}, we have the following identification:
\begin{align*}
E[Y(M_2(M_1(0), 1), M_1(0), 0)]=\iint h_1(w,x)p(w \mid A=0, x)p(x)dwdx.
\end{align*}
\end{theorem}

\begin{remark}
In contrast to traditional proximal causal inference frameworks, the identification formula in Theorem~\ref{thm: Identification 1} admits a second, mathematically equivalent representation. By the law of iterated expectations, the right-hand side can be equivalently expressed as
\begin{equation*}
E\left[ \frac{I(A=0)}{P(A=0 \mid X)} h_1(W,X) \right].
\end{equation*}
This alternative formulation completely avoids the estimation of the conditional density of the proxy $W$, shifting instead to the estimation of the treatment propensity score $P(A=0 \mid X)$ alongside the bridge function $h_1(W,X)$.
\end{remark}

\subsection{Treatment-model-based identification and hybrid identification}
In this subsection, we establish two alternative proximal identification methods. Similarly, we require another completeness assumption along with bridge functions that satisfy another set of specific integral equations.
\begin{assumption}[Completeness of $W$ for $M_1$]
\label{asm:completeness2}
\leavevmode 
\begin{enumerate}[label=(\roman*), itemsep=0.5ex, topsep=0.5ex]
    \item For any square-integrable function $g$ and any value $x$, if
    \[
    E\big[ g(M_1) \mid W, A = 1, X = x \big] = 0 \quad \text{almost surely},
    \]
    then $g(M_1) = 0$ almost surely.
    \item For any square-integrable function $g$ and any value $x$, $m_2$, if
    \[
    E\big[ g(M_1) \mid W, M_2 = m_2, A = 0, X = x \big] = 0 \quad \text{almost surely},
    \]
    then $g(M_1) = 0$ almost surely.
\end{enumerate}
\end{assumption}

\begin{assumption}[Bridge Function q]
\label{asm: bridge function 2}
Suppose that there exist confounding bridge functions $q_1(Z,X)$ and
$q_0(Z,M_2,X)$ that almost surely satisfy
\begin{align}
&E\left[q_1(Z,X) \mid W,A=1,X\right]=\frac{p(W \mid A=0,X)}{p(W \mid A=1,X)},
\label{eq:formula5}\\
&E\left[q_1(Z,X) \mid W, M_2, A=1, X\right]\frac{P(A=1 \mid W,M_2,X)}{P(A=0 \mid W,M_2,X)}=E\left[q_0(Z,M_2,X) \mid W, M_2, A=0, X\right].
\label{eq:formula6}
\end{align}
\end{assumption}

Assumptions~\ref{asm:completeness2} and \ref{asm: bridge function 2} establish the theoretical foundation for the treatment bridge function. Assumption~\ref{asm:completeness2} serves as an alternative completeness condition, emphasizing that the proxy $W$ possesses sufficient variability to represent the unmeasured recanting witness. The existence of treatment bridge functions in Assumption~\ref{asm: bridge function 2} follows a mathematical rationale analogous to our earlier discussion on the outcome bridge functions.

It is worth noting that we do not directly use the two integral equations from Assumption~\ref{asm: bridge function 2} for estimation; instead, we construct two equivalent formulations of these integral equations.

\begin{proposition}
\label{prop:qestiamtion}
The functions $q_1$ and $q_0$ solve the integral Equations~\eqref{eq:formula5} and \eqref{eq:formula6} if and only if they satisfy the following conditional expectation Equations~\eqref{eq:formulapro1} and \eqref{eq:formulapro2}, respectively:
\begin{align}
&E\left[\frac{I(A=1)}{P(A=1 \mid X)}q_1(Z,X)-\frac{I(A=0)}{P(A=0 \mid X)} \middle|  W,X\right]=0,
\label{eq:formulapro1}\\
&E\left[I(A=1)q_1(Z,X)-I(A=0)q_0(Z,M_2,X) \mid W,M_2,X\right]=0.
\label{eq:formulapro2}
\end{align}
\end{proposition} 

Similar to Proposition~\ref{prop: bridge function 1}, the following proposition connects the treatment-model-based bridge functions directly to the true distribution of the hidden recanting witness $M_1$.

\begin{proposition}
\label{prop:bridge 2}
By Assumptions~\ref{asm:positivity}-\ref{asm:Cross-world Independence}, \ref{asm:proxies}, \ref{asm:completeness2}, and \ref{asm: bridge function 2}, we have the following two equations hold almost surely:
\begin{align}
&E\left[q_1(Z,X) \mid M_1,A=1,X\right]=\frac{p(M_1 \mid A=0,X)}{p(M_1 \mid A=1,X)},
\label{eq:formula7}\\
&E\left[q_1(Z,X) \mid M_2, M_1, A=1, X\right]\frac{P(A=1 \mid M_2, M_1, X)}{P(A=0 \mid M_2, M_1, X)}=E\left[q_0(Z,M_2,X) \mid M_2, M_1, A=0, X\right].
\label{eq:formula8}
\end{align}
\end{proposition}

Based on Propositions~\ref{prop: bridge function 1} and \ref{prop:bridge 2}, we obtain the following two identification methods: the Proximal Hybrid
Estimation (PHE), and the Proximal Inverse Probability Weighting (PIPW).

\begin{theorem}[PHE of $\psi$]
\label{thm:identify2}
Under Assumptions~\ref{asm:positivity}-\ref{asm:Cross-world Independence}, \ref{asm:proxies}, \ref{asm:completeness}(ii), \ref{asm:completeness2}(i), and further assuming the existence of bridge functions defined by Equation \eqref{eq:formulah1} in Assumption~\ref{asm: bridge function 1} and Equation \eqref{eq:formula5} in Assumption~\ref{asm: bridge function 2}, we have the following identification:
$$E\left[Y(M_2(M_1(0), 1), M_1(0), 0)\right]=E\left[\frac{I(A=1)}{P(A=1 \mid X)}h_0(W,M_2,X)q_1(Z,X)\right].$$
\end{theorem}

\begin{theorem}[PIPW of $\psi$]
\label{thm:identify3}
Under Assumptions~\ref{asm:positivity}-\ref{asm:Cross-world Independence} and Assumptions~\ref{asm:proxies}, \ref{asm:completeness2} and \ref{asm: bridge function 2}, we have the following identification:
$$E[Y(M_2(M_1(0), 1), M_1(0), 0)]=E\left[\frac{I(A=0)}{P(A=1 \mid X)}Yq_0(Z,M_2,X)\right].$$
\end{theorem}
Thus, we have obtained three identifications, each relying on two bridge functions: POR depends on $h_0$ and $h_1$, PHE on $h_0$ and $q_1$, and PIPW on $q_1$ and $q_0$. 

\section{Semiparametric inference}
\label{sec:Semiparametric inference}
\subsection{The semiparametric efficiency bound}
In this section, we consider inference for the parameter $\psi$ under the semiparametric model $\mathcal{M}_{sp}$ that places no restrictions on the observed data other than Assumption~\ref{asm: bridge function 1}. As previously established, $\psi$ admits three distinct identification strategies. However, direct plug-in estimators based on these identification functionals suffer from first-order bias, meaning their bias decays at the same slow rate as the nonparametric estimation error of the nuisance functions. To overcome this and achieve $\sqrt{n}$-consistency, we propose a Proximal Multiply Robust (PMR) identification based on the efficient influence function of $\psi$ \citep{Robins2008, Robins2017}. This orthogonalization approach eliminates the first-order bias, reducing the impact of nuisance estimation errors to second-order remainder terms. One immediate benefit of this approach is that the resulting estimator possesses the multiply robust property. Furthermore, provided that the nonparametric estimators converge to their respective true functions at sufficiently fast rates, the PMR estimator achieves asymptotic normality and local semiparametric efficiency \citep{newey1990semiparametric, bickel1993efficient}. We need the following assumption to derive the efficient influence function.

\begin{assumption}[Regularity Conditions]\label{asm: Regularity conditions}
Consider the following conditional expectation operators. We assume that:
\begin{enumerate}
    \item[(i)] For the operator $T_0: L^2(W,M_2,X) \to L^2(Z,M_2,A=0,X)$ defined as
    $$ T_0(g) \coloneqq  E\!\left[ g(W,M_2,X)\mid Z, M_2, A=0, X \right], $$
    at the true data-generating mechanism, $T_0$ is surjective.

    \item[(ii)] For the operator $T_1: L^2(W,M_2,X) \to L^2(Z,A=1,X)$ defined as
    $$ T_1(g) \coloneqq E\!\left[ g(W,M_2,X)\mid Z,A=1,X \right], $$
    at the true data-generating mechanism, $T_1$ is surjective.
\end{enumerate}
\end{assumption}

As noted in \cite{Ying2023} and \cite{Dukes2023}, this condition relies on the functions $L_2(W, M_2, X)$ being
rich enough such that any element in $ L_2(Z, M_2, A = 0, X)$ and $L_2(Z, A = 1, X)$ can be
generated via the conditional expectation map. We then obtain the following result.
\begin{theorem}[Efficient Influence Function]
\label{thm:EIF}
Under Assumptions~\ref {asm:positivity}-\ref{asm:completeness}, \ref{asm:completeness2}, and further assume that there exist unique $h_1$ and $h_0$ that solve Equations~\eqref{eq:formulah1} and \eqref{eq:formulah2} at all data laws that belong to $\mathcal{M}_{sp}$. Moreover, suppose that at the true data law there exist unique $q_0$ and $q_1$ that solve Equations~\eqref{eq:formula5} and \eqref{eq:formula6}, and that Assumption~\ref{asm: Regularity conditions} holds. Then we have the efficient influence function:
\begin{align*}
&EIF(O)=\frac{I(A=1)}{P(A=1 \mid X)}q_1(Z,X)[h_0(W,M_2,X)-h_1(W,X)]\\&+\frac{I(A=0)}{P(A=1 \mid X)}q_0(Z,M_2,X)[Y-h_0(W,M_2,X)]
+\frac{I(A=0)}{P(A=0 \mid X)}[h_1(W,X)-\eta(X)]+\eta(X)-\psi,
\end{align*}
where $\eta(X)=E\left[h_1(W,X) \mid A=0,X\right]$ and $O = (Y,Z,W,M_2,A,X)$. Moreover, the corresponding semiparametric local efficiency bound of $\psi$ is $E[EIF(O)^2]$.

\end{theorem}

Based on the influence function obtained in Theorem~\ref{thm:EIF}, the following theorem demonstrates the multiply robust property of the PMR method. Specifically, with a superscript $*$ attached to the nuisance functions to denote their probability limit, we partition the model space $\mathcal{M}_{sp}$ into four subsets, and the consistent result can be obtained as long as at least one of them is correctly specified.

\begin{theorem}
\label{thm:mr}
Under Assumptions~\ref{asm:positivity}-\ref{asm: bridge function 2}, if at least one of the following models is correctly specified:
\begin{align*}
\mathcal{M}_1: & \quad P^{*}(A=1\mid X)\ \text{and}\ \{ h^{*}_0, h^{*}_1\}\ \text{are correctly specified};\\
\mathcal{M}_2: & \quad P^{*}(A=1\mid X)\ \text{and}\ \{ q^{*}_0, q^{*}_1\}\ \text{are correctly specified};\\
\mathcal{M}_3: & \quad P^{*}(A=1\mid X)\ \text{and}\ \{ h^{*}_0, q^{*}_1\}\ \text{are correctly specified};\\
\mathcal{M}_4: & \quad p^{*}(W\mid A=0, X)\ \text{and}\ \{ h^{*}_0, h^{*}_1\}\ \text{are correctly specified};
\end{align*}
we have
\begin{align*}
&\psi=E\bigg[\frac{I(A=1)}{{P^{*}}(A=1 \mid X)}q^{*}_{1}(Z,X)[{h}^{*}_0(W,M_2,X)-h^{*}_1(W,X)]+\\&\frac{I(A=0)}{P^{*}(A=1 \mid X)}q^{*}_0(Z, M_2, X)[Y-h^{*}_0(W,M_2,X)]+\frac{I(A=0)}{P^{*}(A=0 \mid X)}[h^{*}_1(W,X)-{\eta^{*}}(X)]+{\eta^{*}}(X)\bigg],
\end{align*}
where $\eta^{*}(X)=\int h^{*}_1(w,X) p^{*}(w\mid A=0, X)dw$.
\end{theorem}


\subsection{Debiased machine learning}

Our next objective is to estimate the nuisance functions via a minimax optimization-based method, and subsequently estimate the parameter of interest using cross-fitting techniques. Specifically, $P(A=1 \mid X)$ can be estimated via random forests, and $\eta(X)$ can be estimated using kernel ridge regression after obtaining the estimated bridge function $h_1$. Recall that the nuisance functions $h_0$, $h_1$, $q_1$, and $q_0$ are solutions to the integral equations; they cannot be estimated by simple standard regressions. However, a nonparametric estimation method based on the reproducing kernel Hilbert space \citep{ Ghassami2022, Ghassami2025} can be used to solve such integral equations. Here, we employ the same technique for estimating the bridge functions.

Let $\mathcal{H}_0$, $\mathcal{H}_1$, $\mathcal{Q}_0$, and $\mathcal{Q}_1$ be normed function spaces. Building upon the conditional moment Equations~\eqref{eq:formulah1}, \eqref{eq:formulah2}, and Equations~\eqref{eq:formulapro1}, \eqref{eq:formulapro2}, we propose the following minimax optimization-based estimators for the bridge functions $h_0$, $h_1$, and $q_1$, $q_0$.
{\small
\begin{align*}
\hat{h}_{0} &= \arg\min_{h \in \mathcal{H}_0} \sup_{f \in \mathcal{Q}_0} E_{n_0} \left[ \{ Y - h(W, M_2, X) \} f(Z, M_2, X) - f^2(Z, M_2, X) \right] - \lambda_{\mathcal{F}}^h \|f\|_{\mathcal{F}}^2 + \lambda_{\mathcal{H}}^h \|h\|_{\mathcal{H}}^2; \\
\hat{h}_{1} &= \arg\min_{h \in \mathcal{H}_1} \sup_{f \in \mathcal{Q}_1} E_{n_1} \left[ \{ h(W, X) - \hat{h}_0(W, M_2, X) \} f(Z, X) - f^2(Z, X) \right] - \lambda_{\mathcal{F}}^h \|f\|_{\mathcal{F}}^2 + \lambda_{\mathcal{H}}^h \|h\|_{\mathcal{H}}^2; \\
\hat{q}_1 &= \arg\min_{q \in \mathcal{Q}_1} \sup_{f \in \mathcal{H}_1} E_n \left[ \left\{ \frac{I(A = 1)}{\hat{P}(A = 1 \mid X)} q(Z, X) - \frac{I(A = 0)}{\hat{P}(A = 0 \mid X)} \right\} f(W, X) - f^2(W, X) \right] \\
&\quad - \lambda_{\mathcal{F}}^q \|f\|_{\mathcal{F}}^2 + \lambda_{\mathcal{Q}}^q \|q\|_{\mathcal{Q}}^2; \\
\hat{q}_0 &= \arg\min_{q \in \mathcal{Q}_0} \sup_{f \in \mathcal{H}_0} E_n \left[ \left\{ I(A = 1)\hat{q}_1(Z, X) - I(A = 0)q(Z,M_2,X) \right\} f(W, M_2, X) - f^2(W, M_2, X) \right] \\
&\quad - \lambda_{\mathcal{F}}^q \|f\|_{\mathcal{F}}^2 + \lambda_{\mathcal{Q}}^q \|q\|_{\mathcal{Q}}^2.
\end{align*}}Here, $E_n[\cdot]$, $E_{n_0}[\cdot]$, and $E_{n_1}[\cdot]$ denote the empirical expectations over the entire dataset and over the subset with $A = 0$ and $A = 1$, respectively. See the Supplementary Material for details of the minimax optimization. 

After obtaining $\hat{h}_0$, $\hat{h}_1$, $\hat{q}_1$, $\hat{q}_0$, $\hat\eta$, and $\hat{P}(A=1 \mid X)$, we have 
{\small
\begin{align*}
\hat{\psi}_{POR} &= E_{n}[\hat{\eta}(X)]; \\
\hat{\psi}_{PHE} &= E_{n}\left[\frac{I(A=1)}{\hat{P}(A=1 \mid X)}\hat{h}_0(W,M_2,X)\hat{q}_1(Z,X)\right]; \\
\hat{\psi}_{PIPW} &= E_{n}\left[ \frac{I(A=0)}{\hat{P}(A=1 \mid X)}Y\hat{q}_0(Z,M_2,X)\right]; \\
\hat{\psi}_{PMR} &= E_{n}\Bigg[ \frac{I(A=1)}{\hat{P}(A=1 \mid X)}\hat{q}_1(Z,X)\{\hat{h}_0(W,M_2,X)-\hat{h}_1(W,X)\} \\
&\quad + \frac{I(A=0)}{\hat{P}(A=1 \mid X)}\hat{q}_0(Z,M_2,X)\{Y-\hat{h}_0(W,M_2,X)\} + \frac{I(A=0)}{\hat{P}(A=0 \mid X)}\{\hat{h}_1(W,X)-\hat{\eta}(X)\}+\hat{\eta}(X) \Bigg].
\end{align*}}

Within this framework, cross-fitting \citep{schick1986asymptotically, Chernozhukov2018} is employed, which separates the estimation of nuisance functions from the estimation of the parameter of interest through the following procedure.

\begin{enumerate}
    \item[(i)] Partition the data into \(K\) equally-sized folds \(\{I_1,\cdots,I_K\}\). 
    \item[(ii)] For each \(\ell \in \{1,\cdots, K\}\), estimate the nuisance functions using data from all folds except \(I_\ell\). 
    \item[(iii)]  For each \(\ell\), compute the empirical expectation of these moment functions using only data from fold \(I_\ell\), yielding \(K\) separate estimates \(\hat{\psi}^{(\ell)}\).
    \item[(iv)] The final estimator is the average of the fold-specific estimates:
$\hat{\psi} = \frac{1}{K}\sum_{\ell=1}^{K} \hat{\psi}^{(\ell)}$.
\end{enumerate}

Next, we present an analysis of convergence rates \citep{Dikkala2020, Kallus2021, Ghassami2022} that serves as the basis for our proposed learning procedure. We assume that all positivity and boundedness conditions hold with some universal constants.
\begin{theorem}[Cross-fitting]
\label{thm:cross-fitting}
Under Assumptions~\ref{asm:positivity}-\ref{asm: bridge function 2}, and assuming the boundedness of $Y$ and all bridge functions, suppose that
{\small
\begin{align*}
&E\left[\left|\frac{\hat q_1(Z,X)}{\hat P(A=1 \mid X)}-\frac{ q_1(Z,X)}{P(A=1 \mid X)}\right|^2\right] =o(a_{1n}^2),
\quad
&&E\left[\left|\frac{\hat q_0(Z,M_2,X)}{\hat P(A=1 \mid X)}-\frac{ q_0(Z,M_2,X)}{P(A=1 \mid X)}\right|^2\right] = o(a_{0n}^2),\\
&E\left[\left|\hat h_1(W,X)-h_1(W,X)\right|^2\right]= o(b_{1n}^2),\quad
&&E\left[\left| \hat h_0(W, M_2, X) - h_0(W, M_2, X)\right|^2\right]= o(b_{0n}^2),\\
&E\bigl[|\hat P(A=1 \mid X)- P(A=1 \mid X)|^2\bigr] = o(c_n^2), \quad
&&E\bigl[|\hat \eta(X)-\eta(X)|^2\bigr] = o(d_n^2).
\end{align*}}We have
\[
\hat\psi - \tilde\psi
=
o_p(\max\{a_{1n}b_{0n},\; a_{1n}b_{1n},\; a_{0n}b_{0n},\; c_{n}b_{1n},\; c_{n}d_{n},n^{-\frac{1}{2}}\gamma_n\}), 
\]
where $\gamma_n=\max\{a_{1n},a_{0n},b_{1n},b_{0n},c_{n},d_{n}\}$, and  $\tilde{\psi}$ is the oracle estimator estimated using the true nuisance functions.

Furthermore, $\hat\psi$  has an
asymptotically normal distribution and attains a $1/\sqrt{n}$ rate of
convergence, provided that the nuisance components are learned using cross-fitting, and achieve
fourth-root rates of convergence in root mean squared error. Moreover, the asymptotic variance of \(\hat{\psi}\) achieves the semiparametric efficiency bound. Specifically,
\[
\sqrt{n}\bigl(\hat{\psi} - \psi\bigr) \;\stackrel{d}{\longrightarrow}\; \mathcal{N}\!\Bigl(0,\;E\bigl[EIF(O)^2\bigr]\Bigr),
\]
where $EIF(O)$ is the efficient influence function characterized in Theorem~\ref{thm:EIF}.  
\end{theorem}

The established asymptotic normality in Theorem~\ref{thm:cross-fitting} enables the construction of valid confidence intervals for $\psi$, with details provided in the Supplementary Material. 

\begin{remark}
We make two remarks regarding the convergence rate requirements of the nuisance estimators. First, the standard $o(n^{-1/4})$ rate is sufficient but not strictly necessary; $\sqrt{n}$-consistency can be achieved through asymmetric rate compensation, provided the product of paired convergence rates, e.g., $a_{1n}b_{1n}$, is $o(n^{-1/2})$. Second, the term $n^{-1/2}\gamma_n$ is asymptotically negligible. Because nonparametric rates are generally slower than $O(n^{-1/2})$, this term is therefore dominated by the cross-product bias terms.
\end{remark}

\section{Simulation}
\label{sec:simulation}
\subsection{Simulation setup}

The data-generating process follows a recursive structural equation model with additive Gaussian noise. The baseline covariates are drawn from a standard multivariate normal distribution, $X \sim \mathcal{N}(0, I_{d_x})$. The binary treatment is generated as $A \sim \text{Bernoulli}(\sigma(X^\top \beta_{xa} + \epsilon_a))$, where $\sigma(\cdot)$ denotes the sigmoid function. All subsequent continuous variables are simulated sequentially as linear combinations of their respective causal ancestors. Specifically, the unobserved recanting witness $M_1$ is generated from $\{A, X\}$, the subsequent mediator $M_2$ from $\{M_1, A, X\}$, the proxy $Z$ from $\{M_1, A, X\}$, and the proxy $W$ from $\{M_1, X\}$. Finally, the continuous outcome $Y$ is generated as a linear function of $\{W, M_2, M_1, A, X\}$. In each of these structural equations, the generative model incorporates corresponding coefficient matrices and mutually independent, zero-mean Gaussian error terms, with variances scaled by identity matrices matching their respective dimensions. Our target is to estimate $\psi$, the true value of which can be approximated numerically via Monte Carlo simulation.

We consider two configurations to evaluate estimator performance with varying $d_x, d_z, d_w$.
In Case 1, $d_x = 3, d_z = 2, d_w = 2$; in Case 2, $d_x = 5, d_z = 3, d_w = 3$; in both cases,  $d_{m_1} = 1, d_{m_2} = 1$.
For each case, we generate datasets with sample sizes $n \in \{200, 500, 1000\}$ and 300 replications. For each simulation replication, the coefficient matrices and vectors are independently drawn from a specific uniform distribution. By resampling the structural coefficients in each replication, we evaluate the estimator's performance across a broad spectrum of underlying causal mechanisms rather than relying on a single fixed parameterization. 

For $n=1000$, we also compare our proposed method with three naive approaches that simply substitute proxy variables for the unmeasured $M_1$:
$Z\text{-as-}M_1$ treats $Z$ as the true $M_1$; $W\text{-as-}M_1$ treats $W$ as the true $M_1$; $ZW\text{-as-}M_1$ treats both $Z$ and $W$ as the true $M_1$.
These naive methods represent common but incorrect practices when analysts have only proxy measurements. 
Additionally, we compute the 95\% confidence intervals using the efficient influence function and evaluate their length and coverage probability.
The Python code to replicate the numerical results in our paper is available at \url{https://github.com/SihanWu03/Proximal-mediation-analysis-with-hidden-recanting-witnesses}.

\subsection{Simulation results}
\begin{figure}[htbp]
    \centering
    \begin{subfigure}{0.5\textwidth}
        \centering
        \includegraphics[width=\textwidth]{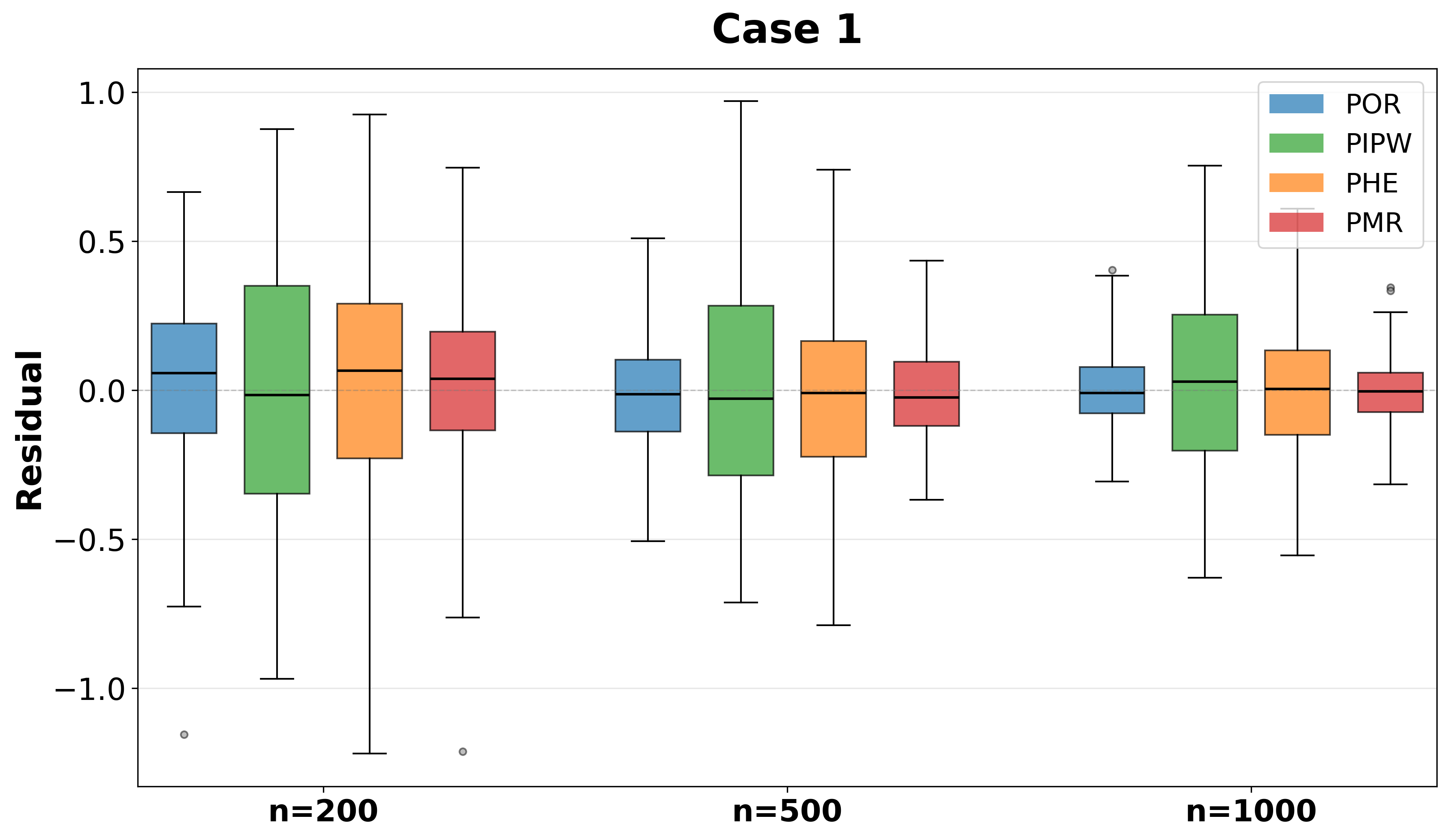}
        \caption{Residual plot for Case 1;}
        \label{fig:case1residuals}
    \end{subfigure}\hfill 
    \begin{subfigure}{0.5\textwidth}
        \centering
        \includegraphics[width=\textwidth]{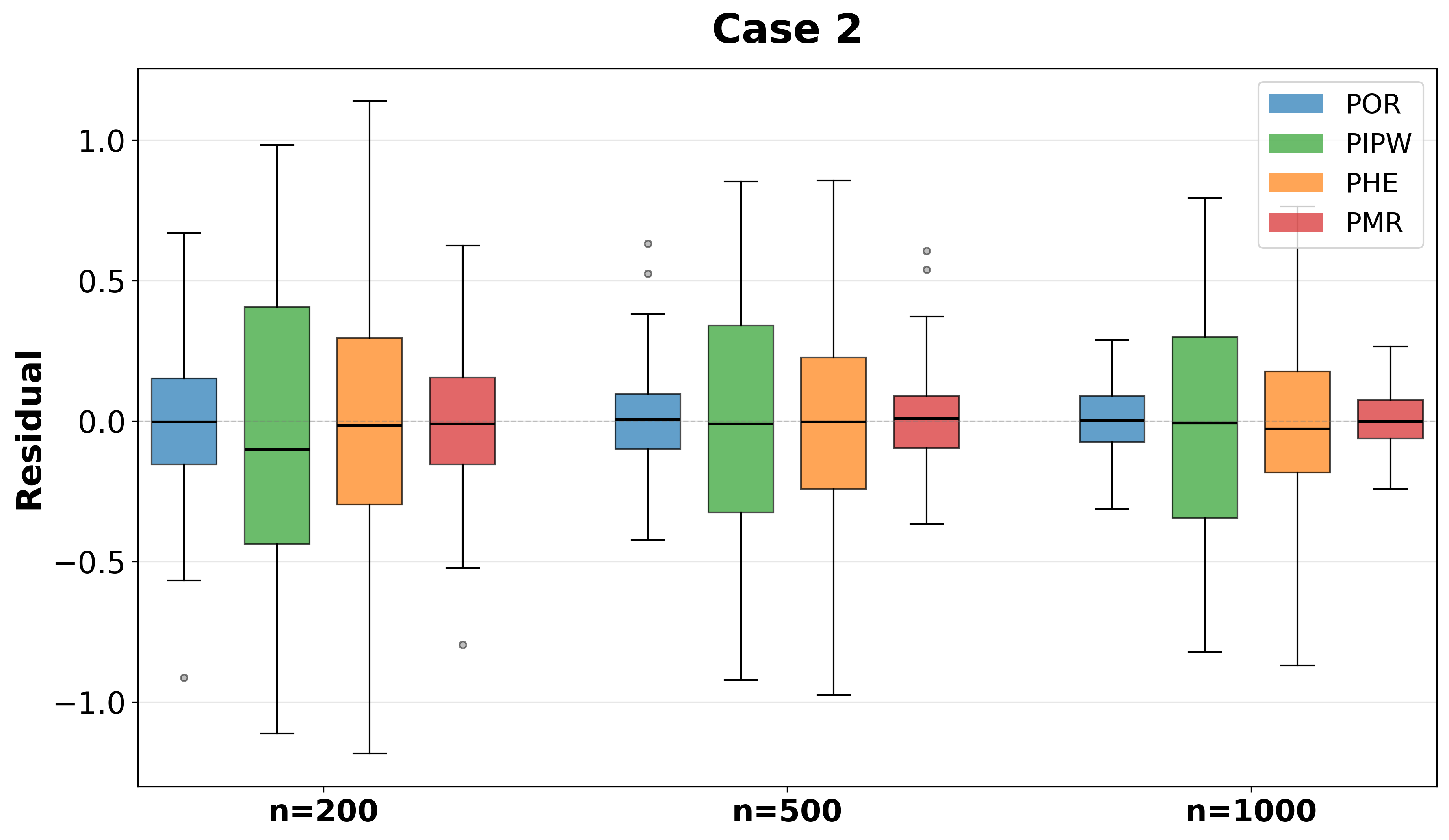}
        \caption{Residual plot for Case 2.}
        \label{fig:case2residuals}
    \end{subfigure}

    \caption{Residual distributions of four estimators (POR, PIPW, PHE, PMR) across sample sizes $n=200, 500, 1000$ for Cases 1 and 2.}

    \label{fig:combined_residuals}
\end{figure}

Figures~\ref{fig:case1residuals} and \ref{fig:case2residuals} display the residual distributions for Cases 1 and 2, respectively. Across all sample sizes, the proposed PMR estimator demonstrates the best performance. As the sample size increases, all estimators exhibit reduced variability, with PMR converging most rapidly toward the true parameter value.

Table~\ref{tab:estimator_results} summarizes the finite-sample performance of the four estimators across two simulation cases. As expected, all methods exhibit decreasing variance, Mean Squared Error (MSE), and Mean Absolute Error (MAE) as the sample size increases, confirming their desirable asymptotic properties. Notably, the proposed PMR estimator consistently achieves superior performance across all scenarios.

\begin{table}[htbp]
\centering
\caption{Simulation results of the proposed four estimators across two cases}
\label{tab:estimator_results}
\begin{tabular}{ll|cccc|cccc}
\hline
& & \multicolumn{4}{c|}{Case 1} & \multicolumn{4}{c}{Case 2} \\
\cline{3-6} \cline{7-10}
$n$ & Method & Bias & Var & MSE & MAE & Bias & Var & MSE & MAE \\
\hline
$200$  & POR  & $0.0263$ & $0.0725$ & $0.0730$ & $0.2179$ & $-0.0057$ & $0.0595$ & $0.0593$ & $0.1907$ \\
     & PIPW & $-0.0074$ & $0.1782$ & $0.1777$ & $0.3668$ & $-0.0187$ & $0.2402$ & $0.2397$ &  $0.4428$ \\
     & PHE  & $0.0242$ & $0.1473$ & $0.1474$ & $0.3115$ & $-0.0124$ & $0.1587$ & $0.1583$ & $0.3249$ \\
     & PMR  & $0.0292$ & $0.0675$ & $0.0681$ & $0.2052$ & $-0.0036$ & $0.0530$ & $0.0528$ & $0.1820$ \\
\hline
$500$  & POR  & $-0.0065$ & $0.0328$ & $0.0327$ & $0.1460$ & $0.0008$ & $0.0230$ & $0.0229$ & $0.1191$ \\
     & PIPW & $-0.0059$ & $0.1244$ & $0.1240$ & $0.3021$ & $-0.0006$ & $0.1533$ & $0.1527$ & $0.3466$ \\
     & PHE  & $-0.0202$ & $0.0778$ & $0.0779$ & $0.2254$ & $-0.0084$ & $0.1023$ & $0.1020$ & $0.2612$ \\
     & PMR  & $-0.0037$ & $0.0245$ & $0.0244$ & $0.1264$ & $0.0015$ & $0.0194$ & $0.0193$ & $0.1096$ \\
\hline
$1000$ & POR  & $-0.0023$ & $0.0146$ & $0.0146$ & $0.0949$ & $0.0051$ & $0.0120$ & $0.0120$ & $0.0904$ \\
     & PIPW & $0.0227$ & $0.0808$ & $0.0810$ & $0.2391$ & $-0.0047$ & $0.1334$ & $0.1330$ & $0.3241$ \\
     & PHE  & $0.0019$ & $0.0474$ & $0.0472$ & $0.1741$ & $-0.0074$ & $0.0745$ & $0.0744$ & $0.2169$ \\
     & PMR  & $-0.0040$ & $0.0105$ & $0.0104$ & $0.0809$ & $0.0051$ & $0.0098$ & $0.0098$ & $0.0802$ \\
\hline
\end{tabular}
\end{table}

Table~\ref{tab:pmr_comparison_long} compares the performance of the proposed PMR estimator against naive approaches. The results demonstrate that naive methods suffer from severe under-coverage and heavy estimation errors, fundamentally invalidating subsequent statistical inferences. In contrast, the PMR framework establishes inferential validity by achieving coverage probabilities close to the nominal level. Furthermore, PMR maintains superior statistical efficiency, consistently yielding strictly narrower confidence intervals and substantially lower mean squared errors.

\begin{table}[htbp]
\centering
\caption{Comparison of the proposed PMR estimator with naive methods ($n=1000$)}
\label{tab:pmr_comparison_long}
\begin{tabular}{llccccc}
\hline
Case & Method & Bias & Var & MSE & CI Length & Coverage \\
\hline
Case 1     & PMR    & $-0.0040$ & $0.0105$ & $0.0104$ & $0.3600$ & $93.0$\% \\
           & $Z\text{-as-}M_1$  & $-0.0076$ & $0.2068$ & $0.2062$ & $0.4191$ & $31.0$\% \\
           & $W\text{-as-}M_1$  & $ 0.0277$ & $0.1305$ & $0.1308$ & $0.3916$ & $33.3$\% \\
           & $ZW\text{-as-}M_1$ & $-0.0326$ & $0.1454$ & $0.1460$ & $0.3768$ & $24.3$\% \\
\hline
Case 2     & PMR  & $0.0051$ & $0.0098$ & $0.0098$ & $0.3376$ & $91.7$\% \\
           & $Z\text{-as-}M_1$  & $-0.0063$ & $0.2115$ & $0.2108$ & $0.3779$ & $20.7$\% \\
           & $W\text{-as-}M_1 $ & $0.0155$ & $0.1143$ & $0.1142$ & $0.3920$ & $33.7$\% \\
           & $ZW\text{-as-}M_1$ & $-0.0046$ & $0.1427$ & $0.1422$ & $0.3940$ & $22.7$\% \\
\hline
\end{tabular}
\end{table}

\section{Real data application}
\label{sec:real_data}
In this section, we apply the proposed PMR framework to data from the National Longitudinal Survey of Youth 1997 \citep{bls_nlsy97} to investigate the wage returns of high school educational tracking. The analytic sample tracks individuals from high school through 2013. The treatment variable $A$ represents the high school track, where $A=1$ indicates a rigorous college preparatory track and $A=0$ represents a general high school track. The outcome $Y$ is the adult hourly wage. The traditional human capital perspective assumes that an intensive curriculum ($A$) increases future wages ($Y$) primarily by boosting a student's cognitive and academic abilities ($M_2$), typically measured by standardized test scores such as the Armed Services Vocational Aptitude Battery.

However, this pure cognitive pathway is heavily confounded by unmeasured non-cognitive traits ($M_1$), such as self-discipline, rule-following, and conformity. Although highly valued by the labor market, these unobserved traits often act as a hidden recanting witness that can destabilize the perceived cognitive mechanism. Because $M_1$ remains unobserved, standard mediation analysis faces an identification challenge.

If one assumes the absence of recanting witnesses, classical approaches \citep{TchetgenTchetgen2012} can be used to estimate the standard NIEs and NDEs:
\begin{align*}
&NIE_0 \coloneqq E[Y(M_2(1), 0)] - E[Y(M_2(0), 0)], \quad &&NDE_0 \coloneqq E[Y(M_2(0), 1)] - E[Y(M_2(0), 0)],\\
&NIE_1 \coloneqq E[Y(M_2(1), 1)] - E[Y(M_2(0), 1)], \quad &&NDE_1 \coloneqq E[Y(M_2(1), 1)] - E[Y(M_2(1), 0)].
\end{align*}

\begin{table}[htbp]
\centering
\caption{Estimated natural direct and indirect effects on hourly wage (\$/hour)}
\label{tab:traditional_estimates}
\begin{tabular}{lccc}
\toprule
\textbf{Estimand} & \textbf{Estimate} & \textbf{95\% CI} & \textbf{Significant} \\
\midrule
$NIE_0$ & $-4.00$ & $[-11.01, 3.01]$ & No \\
$NIE_1$ & $4.18$ & $[1.57, 6.78]$ & Yes \\
\midrule
$NDE_0$ & $-4.65$ & $[-14.28, 4.99]$ & No \\
$NDE_1$ & $3.53$ & $[-1.80, 8.85]$ & No \\
\bottomrule
\end{tabular}
\end{table}

Table~\ref{tab:traditional_estimates} reports these empirical results. The classical estimators exhibit large uncertainty. The potential failure of these traditional methods stems from a violation of the identifying assumptions required by standard mediation estimators: due to the hidden recanting witness $M_1$, both NIEs and NDEs are nonparametrically unidentifiable.

To break this confounding structure, we evaluate two path-specific effects that isolate the path-specific wage contrast mediated through measured cognitive ability:
\begin{align*}
&PSE_0 \coloneqq E[Y(M_2(M_1(0), 1), M_1(0), 0)] - E[Y(0)],\\
&PSE_1 \coloneqq E[Y(1)] - E[Y(M_2(M_1(1), 0), M_1(1), 1)].
\end{align*}
To capture the latent non-cognitive traits ($M_1$), we utilize two distinct sets of behavioral proxies from the NLSY97. The first set, denoted as $Z$, captures school-related engagement behaviors, including tardiness, absenteeism, and homework time. The second set, denoted as $W$, reflects severe behavioral infractions, including school suspensions, physical fighting, and arrest history. Specifically, mild disengagement ($Z$) does not directly depress earnings, nor does academic tracking ($A$) directly deter severe infractions ($W$). Moreover, neither proxy directly affects test scores. Consequently, they serve as candidate proxies for the latent non-cognitive traits ($M_1$).

\begin{table}[htbp]
\centering
\caption{Estimated cognitive path effects on hourly wage (\$/hour)}
\label{tab:estimates}
\begin{tabular}{lccc}
\toprule
\textbf{Estimand} & \textbf{Estimate} & \textbf{95\% CI} & \textbf{Significant} \\
\midrule
$PSE_0$ & $-3.47$ & $[-6.47, -0.46]$ & Yes \\
$PSE_1$ & $3.18$ & $[1.17, 5.20]$ & Yes \\
\bottomrule
\end{tabular}
\end{table}
Table~\ref{tab:estimates} reports the results from our proposed PMR framework. Unlike the unstable and invalid traditional estimators, our method estimates path-specific effects under the maintained assumptions. For the college preparatory track ($A=1$), $PSE_1$ reveals a significant positive wage return of \$3.18 per hour. This indicates that, under academic tracking ($A=1$), the enhancement of cognitive ability induced by academic tracking yields substantial economic returns.

Interestingly, for the general high school track ($A=0$), the PMR estimator suggests a negative path-specific contrast ($PSE_0 = -3.47$). 
Under the maintained proximal identification assumptions, the negative estimate of $PSE_0$ is consistent with a mismatch pattern: cognitive gains induced by college-preparatory tracking may not translate into wage gains when non-cognitive traits remain at the general-track level.
Theoretically, possessing high cognitive ability without complementary non-cognitive skills is associated with heightened occupational friction, job dissatisfaction, workplace conflicts, and higher turnover. Thus, our framework provides evidence that isolated academic gains may yield negative returns if they decouple from robust non-cognitive skill cultivation.

In summary, these empirical findings underscore the necessity of accounting for hidden recanting witnesses. The analysis shows that the joint dynamics of cognitive and non-cognitive skill formation may play an important role in shaping the economic returns of high school tracking. Consequently, the findings suggest that policies focusing only on test scores may overlook non-cognitive channels.

  \section{Discussion}\label{sec:disc}

This paper addresses a fundamental challenge in causal mediation analysis arising from the presence of unobserved recanting witnesses by proposing a novel proximal framework for path-specific effects. We develop three identification strategies based on bridge functions and further construct an efficient and multiply robust estimator. This estimator requires only a relaxed set of conditions for consistency and achieves the semiparametric efficiency bound when all models are correctly specified. Moreover, we propose a debiased machine learning procedure for valid statistical inference.

Several promising avenues remain for future research. First, future studies could explore scenarios where the subsequent mediator is also unobserved. Second, although our discussion focuses on identifying a specific mediation pathway, extending this strategy to evaluate other pathway effects is of great interest. Third, broadening the scope to other estimands, such as population-level effects \citep{hubbard2008population, fulcher2020robust}, presents a practically relevant extension. 
Fourth, the proposed method can be applied to policy learning following \cite{nabi2019learning,bian2026double}.
Finally, as an alternative to proximal causal inference, future research could leverage instrumental variables \citep{imbens1994identification,angrist1996identification} and possibly invalid proxies \citep{yu2025fortified, rakshit2025adaptive} to address unobserved recanting witnesses.

\putbib[bibfile_main] 
\end{bibunit}

\end{document}